\documentclass[11pt]{article}
\usepackage{moriond,epsfig}

\def\Journal#1#2#3#4{{#1} {\bf #2}, #3 (#4)}
\def\PRL{\em Phys. Rev. Lett.}

\begin{document}

\title{Randall-Sundrum Gravitons and Black Holes at the LHC}

\author { K.M. Black}

\address{ Laboratory for Particle Physics and Cosmology, Harvard University,
18 Hammond Street, Cambridge, MA USA}

\maketitle

\abstracts{Models predicting the existence of extra spatial dimensions offer
compelling and novel solutions to outstanding problems of the Standard Model. In such models, our universe exists on 
a 4 dimensional brane embedded in a larger dimensional space time. By allowing gravity to propagate in the bulk  
the gravitational coupling could be comparable with the other gauge interactions thus removing the hierarchy problem. 
The phenomenology of these models could have dramatic observable effects at the LHC including
the production and decay of gravitons and mini black holes. In this note we summarize feasibility studies for the discovery 
of strong gravitational interactions at the LHC.}

\section{Theoretical Motivation}

In the last decade, a number of new approaches to solving the hierarchy problem have been developed. One of the most novel
approaches involves the addition of extra spatial dimensions. The perceived of  weakness gravity is postulated to arise from the fact that the gravity is allowed to propagate into the extra dimensions while the rest of the Standard Model particles are confined to the  standard three spatial dimensions.

 Randall and Sundrum \cite{RS1,RS2} were amongst the first to develop such models. In the original model there
are two 3-dimensional branes embedded in a universe with one extra spatial dimension. The two branes are separated by a distance
in the extra dimension which gravity, but not the rest of the Standard Model particles, can propagate in. 

The model has a space time metric given by:
\begin{equation}
ds^{2} = e^{-2k|z|} [dt^{2} - d{x}^{2}] - dz^{2}
\label{RS-metric}
\end{equation}

where x labels the familiar 3-dimensional space, z labels the  extra spatial dimension, and $k=L^{-1}$ is the 
inverse of the radius of curvature of the extra dimension. Since gravity propagates in the bulk, the effective strength 
of gravity on our brane appears weaker. Using equation \ref{RS-metric} the hierarchy relation can be written in
the form \cite{RS1,RS2}
\begin{equation}
M^{2}_{Pl} = \frac{M^{3}_{5}}{k}(1 - e^{-2 \pi k r})
\label{mass-eff}
\end{equation}

where $M_{Pl}$ is the effective Planck mass on the 3-dimensional brane, $M_{5}$ is the Planck mass in the bulk, and $k$ is the
curvature of the extra dimension. This insight shows how one can remove the hierarchy problem in a rather simple way
by modifying the geometry of space-time. Current experimental limits extracted from direct searches place constraints that vary on the 
coupling and range from several hundred GeV to approximately 1 TeV at the 95\% confidence level.

\section{Collider Phenomenology}

Since the graviton would couple to the energy-momentum tensor, in principle its presence should be observable through
any Standard Model process. However, the most promising discovery mode at hadron colliders involves the direct production
of gravitons and observation of their decay products. For example, a graviton could be produced and decay into an
electron-positron pair and observed as a new heavy resonance in the dielectron mass distribution.

Banks and Fischler pointed out \cite{Willy} the possibility of black hole production in high energy scattering
in models where gravity becomes strong at the TeV scale. Following this work,
Dimopoulos and Landsberg \cite{GREG} developed a model for black hole production at the LHC.. For example, equation \ref{mass-eff} shows that in the Randall-Sundrum
model the bulk Planck mass could be lowered to the electroweak scale which is accessible at the LHC. Dimopoulos and Landsberg proposed
the following (simplified) model of black hole production by colliding partons. 

Following ref \cite{GREG}, in $n$ dimensions the Schwartzchild radius is given by \cite{Myers}:

\begin{equation}
R_{s} = \frac{1}{\sqrt{\pi} M_{Pl}}[ \frac{M_{BH}}{M_{Pl}} (\frac{ 8 \Gamma(\frac{n+3}{2})}{n+2})^{\frac{1}{n+1}}]
\label{bh-eq}
\end{equation}

Two colliding partons with center-of-mass energy  equal to the mass of the black hole $\sqrt{s} = M_{BH}$ will form
a black hole if the impact parameter is smaller than Schwartzchild radius in equation \ref{bh-eq}. The cross-section is thus
estimated from purely geometric considerations and is on the order of 

\begin{equation}
\sigma_{BH} \approx \pi R^{2}_{s}
\end{equation}

Depending on the number of extra dimensions and the Planck mass the black hole cross-section ranges from a few fb to 0.5 nb.
It should be noted that the large cross-sections (compared to other beyond the standard model theories)
come from essentially two factors:
\begin{itemize}
\item The lack of any small coupling constant.
\item The geometric form of the cross-section which lacks any phase space suppression factor. 
\end{itemize}

The black hole is expected to decay via Hawking radiation 'democratically' into all Standard Model particles with equal
probability. The black hole decay signature consists of two or more high energy particles decaying isotropically.

\section{Feasibility Studies}

The ATLAS and CMS experiments have conducted several studies to estimate the sensitivity for the discovery of both Randall-Sundrum gravitons
and black holes. Because of space limitations I focus on only two analysis (for additional studies see refs \cite{cms_diphoton,cms_dijet,atlas_rs,atlas_fab,cms_bh}).

\subsection{Graviton Searches at CMS}

The search  for dilepton resonances is one of the most promising discovery channels for a Randall-Sundrum graviton. The CMS collaboration has undertaken
a detailed feasibility study in this channel ~\cite{cms_dimuon}. In this note , we limit ourselves to the dimuon channel.  The signature is a resonance in the dimuon invariant mass spectrum. The main background for this search is Drell-Yan production of dimuon pairs. Other backgrounds include $t \bar{t}$ and diboson production - however these are negligible compared to the dominant Drell-Yan background.

The study requires events with  two reconstructed muons. In order to simulate the trigger requirements, they require that at least one of the muons has $|\eta| < $ 2.1 and
$p_{T} > $ 24 GeV. The total trigger efficiency depends on the resonance mass but is between 94\%-97\% for gravitons within the mass range from 1 to 5 TeV. 
Since very high momentum muons can radiate photons with a non-negligible probability, reconstructed photons within a $\Delta$ R cone of 0.1 around a muon are 
searched for. If one is found, the four-momentum of the closest one is added to that of the muon improving the mass resolution. 

In order to estimate the discovery potential, ensemble experiments are performed. Pseudo-experiments are formed and the invariant mass spectrum is fit to two hypothesis. First, a signal and background hypothesis and secondly a background only hypothesis are assumed. The logarithm of the ratio of likelihoods of the two fits is used estimate the signal significance. Pseudo-experiments are constructed from both signal and background and background only Monte Carlo. From the mean value of the log-likelihood distribution one can estimate the amount of data needed to reach 5 $\sigma$ discovery.

\begin{figure}

\begin{center}
\includegraphics[height=1.75in]{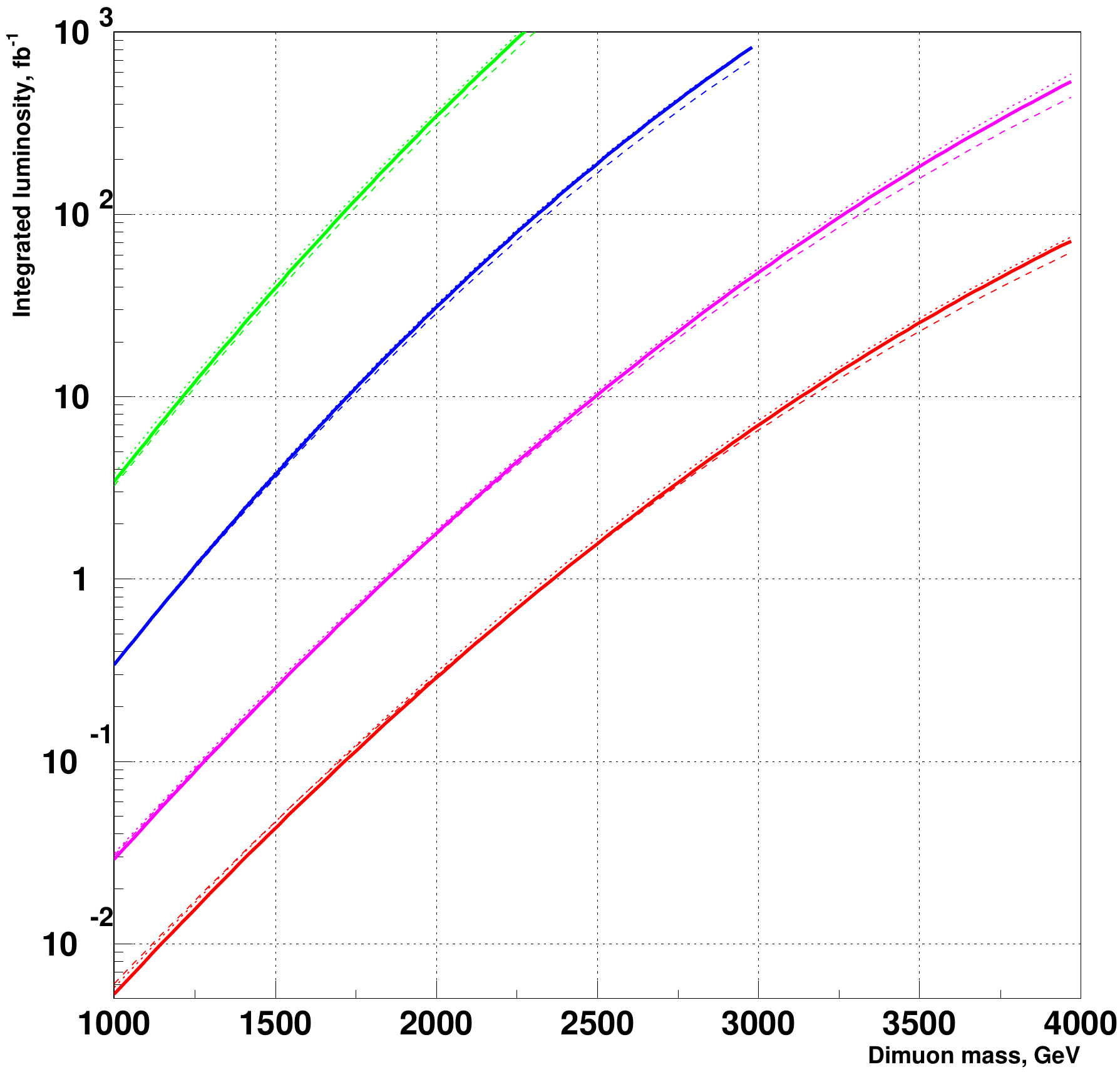} 
\includegraphics[height=1.75in]{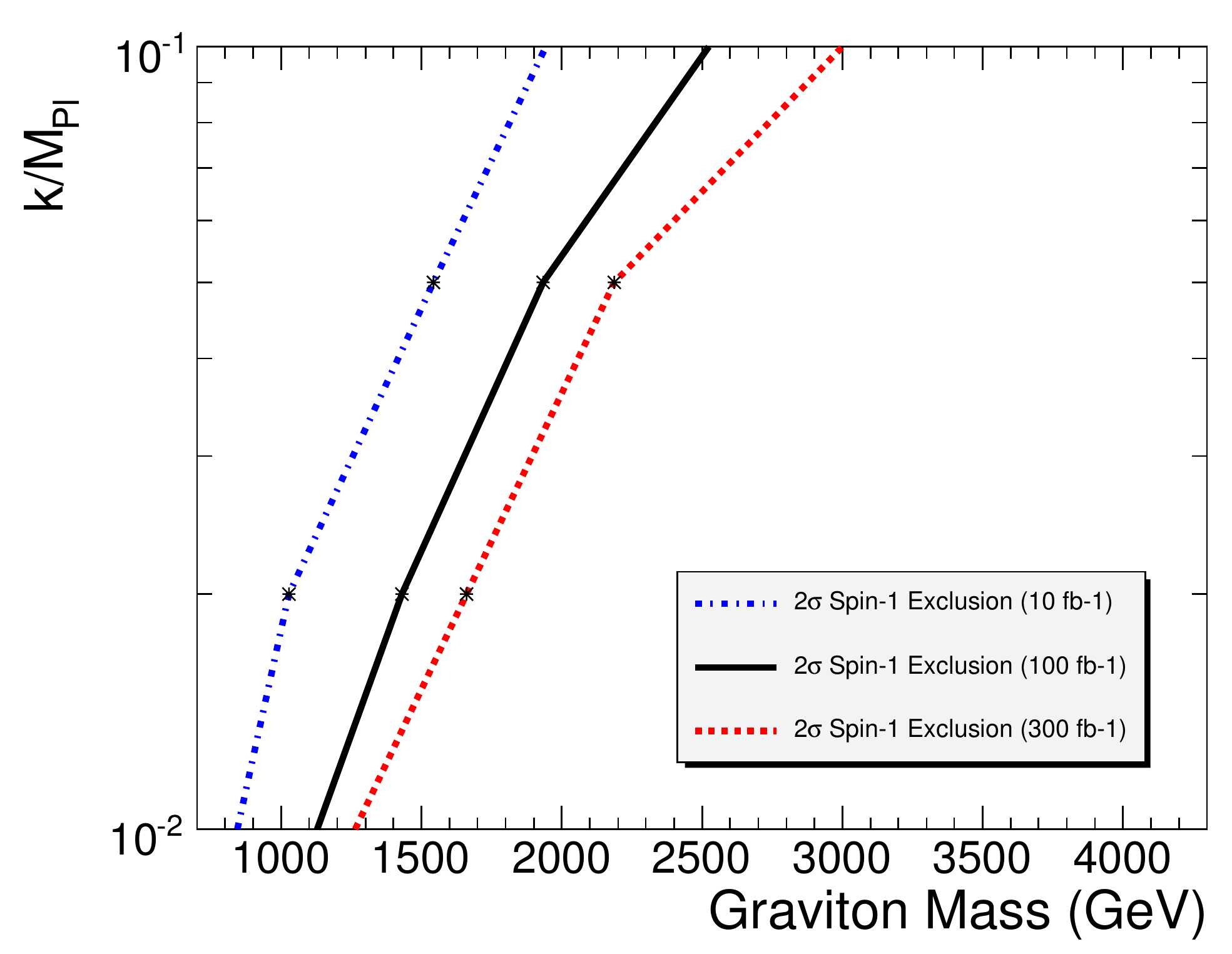} 
\end{center}
\caption{Discovery potential at CMS  in the dimuon channel for Randall-Sundrum gravitons(left) and spin determination potential (right) .}
\label{cms-dimuons-fig}
\end{figure}

 The study also includes evaluation of various systematic 
effects including higher order corrections to the Drell-Yan cross section, parton distribution uncertainties, misalignment and pile-up effects, magnetic field uncertainties, and potential variations in the Drell-Yan shape. The results are shown in figure ~\ref{cms-dimuons-fig} and indicate that the amount of integrated luminosity needed for discovery varies from 10-20 $\rm pb^{-1}$ for small masses and large coupling to several hundred $\rm fb^{-1}$ for large masses and small coupling.

The dilepton final state is completely specified by two variables which which are taken to be  invariant mass and the scattering angle of collision of the dilepton pair. The angular distribution
of the dilepton final state depends on the spin of the resonance. A similar study was undertaken to determine how much data is needed to distinguish between various spins and the results in figure ~\ref{cms-dimuons-fig} show  the ability to separate a spin 1 and spin 2 resonance as a function
of the invariant mass and coupling. 

\section{Black Hole Searches at ATLAS}

The ATLAS collaboration has recently estimated the sensitivity to black hole production.
A signature of black hole production at a hadron collider is the anomalous production of events with many high energy leptons
and jets. The main backgrounds are expected to be $t \bar{t}$ pair production, W/Z + jet production, diphoton + jet production, and multijet production.

Although the signatures are expected to be quite dramatic and the cross-section quite large, a detailed study was undertaken
to optimize the discovery potential. The event selection required events with one high $p_{T}$ electron or muon with $p_{T} > $ 50 GeV,
and the scalar sum of the $p_{T}$ of all electrons, muons, and jets in the event ($\sum{p_{T}}$) $>$ 2.5 TeV. 

\begin{figure}

\begin{center}
\includegraphics[height=1.75in]{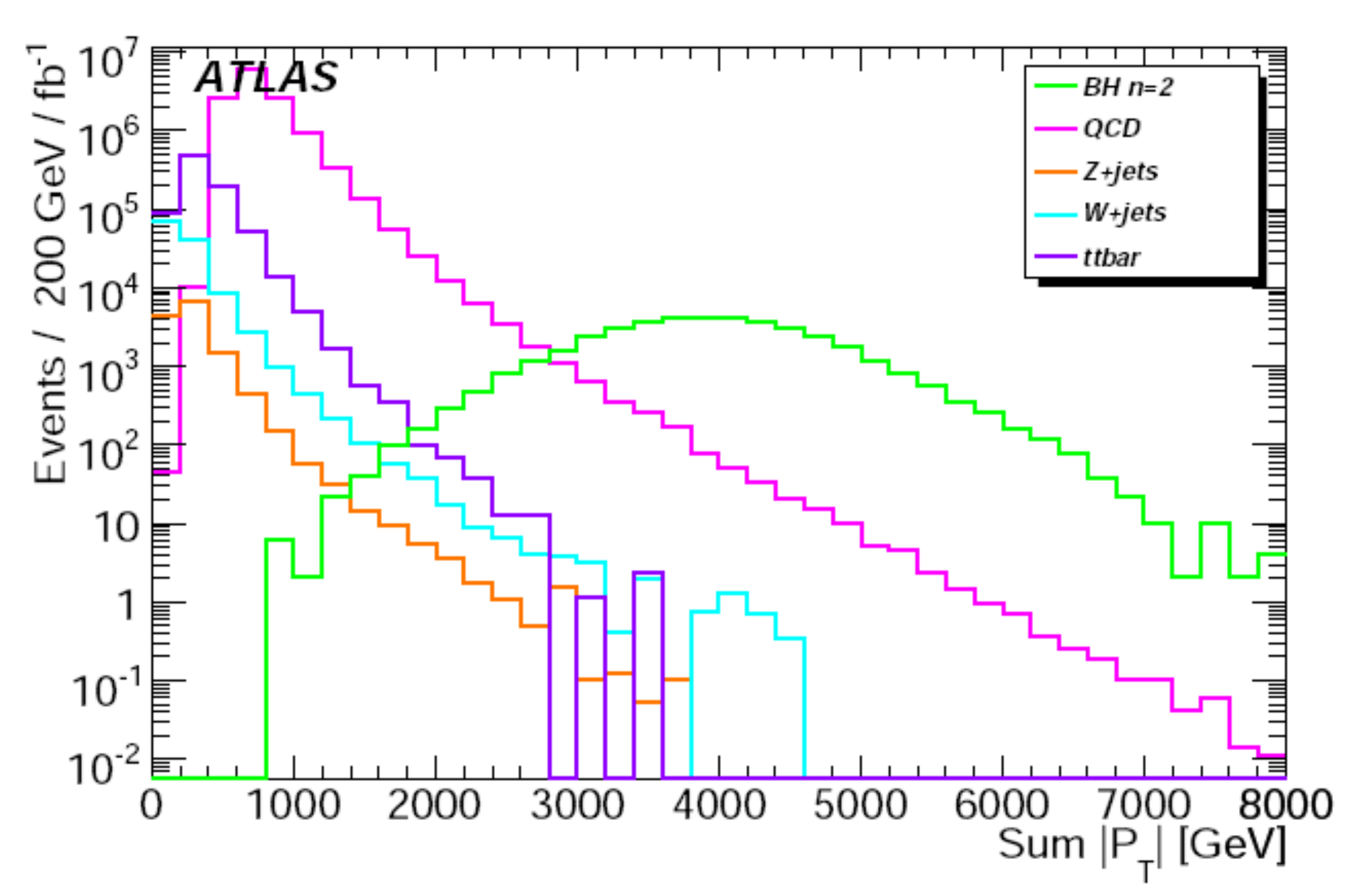} 
\includegraphics[height=1.75in]{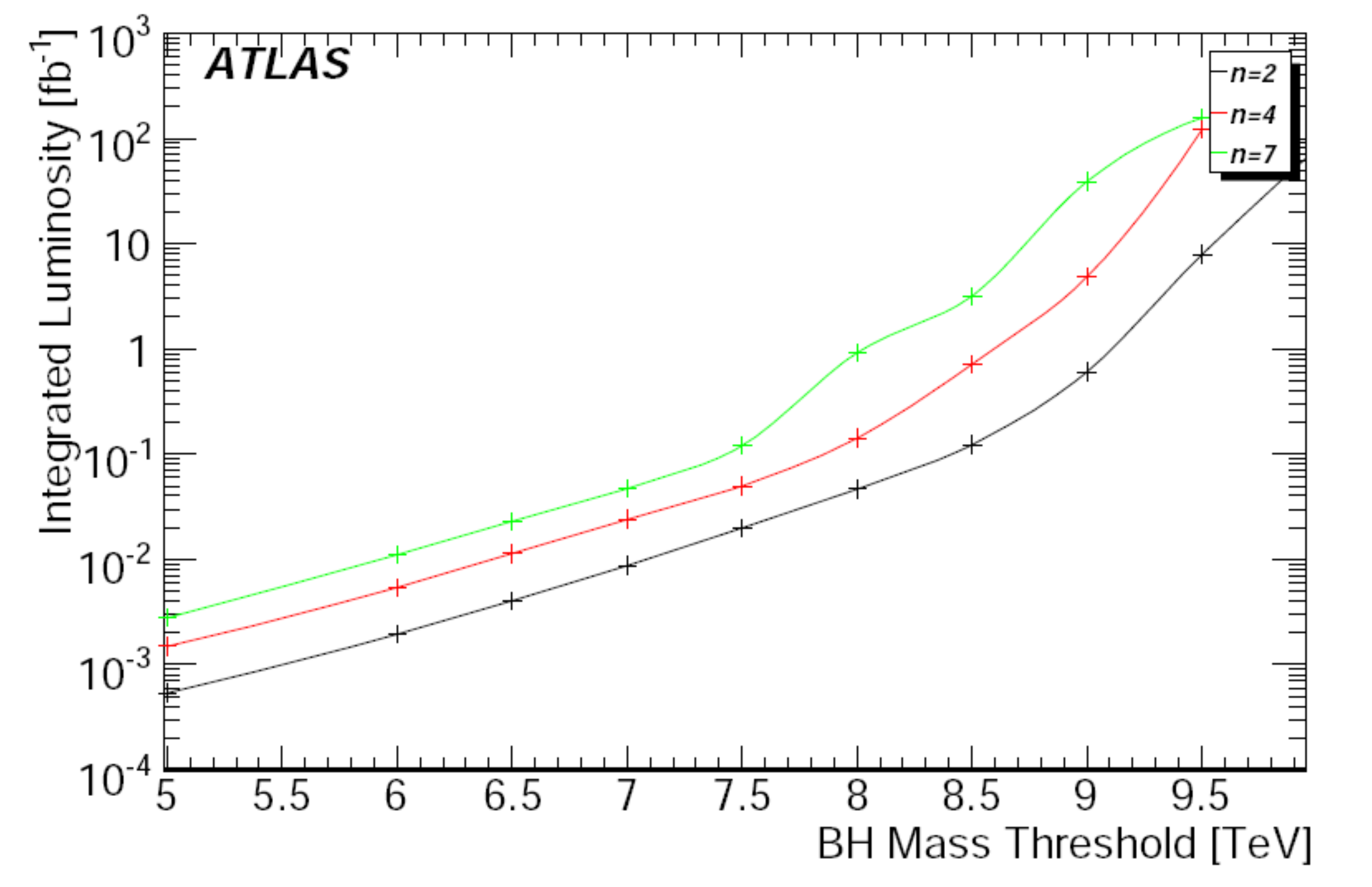} 
\end{center}
\caption{Scalar sum of $p_{T}$ of all objects in the event (left) and black hole discovery potential as a function of black hole mass threshold (right).}
\label{atlas-black}
\end{figure}

The requirement on the high $p_{T}$ lepton is highly efficient against the multijet background while the very selective requirement on
the scalar $p_{T}$ sum of the reconstruction objects was found to be very high rejection to all backgrounds. The total rejection was found to vary from  $10^{5}$ 
for the vector boson + jets to $10^{7}$ for multijet events. The scalar $p_{T}$ sum distribution for signal and background is shown in figure \ref{atlas-black}.
The signal efficiency was found to depend on the exact parameters of the signal (number of extra-dimensions and Planck mass) and varied from about 20\% to
60\% . 

The discovery potential was then estimated by counting events using  $sig=\frac{S}{\sqrt{B}}$  to estimate the significance. The significance 
was required to be greater than 5 with at least 10 signal events. Depending on the
number of extra dimensions and the black hole mass threshold the amount of luminosity required for discovery was found to vary from a few $\rm pb^{-1}$ to several hundred $\rm fb^{-1}$.

\section{Conclusions}

The discovery of extra spatial dimensions, strong gravity ,and black holes at the LHC would be truly fascinating and spectacular. Many detailed studies
to estimate the CMS and ATLAS potential to such possibilities have been undertaken. The amount of integrated luminosity for discovery varies as 
a function of the model parameters but could be as small as a few inverse picobarns. Both collaborations eagerly await the startup of the LHC
later this year to probe the possibility of strong gravity at the TeV scale.

\end{document}